\documentclass[12pt]{article}

\pdfoutput=1
\usepackage{graphicx} 
\usepackage{xspace}
\usepackage{url}
\usepackage{comment}
\usepackage{amsmath}
\usepackage{amssymb}
\usepackage{fancyhdr}  
\usepackage{hepunits}
\usepackage[usenames,dvipsnames]{color}  
\usepackage{enumitem}
\usepackage{multirow}
\usepackage{lineno}
\usepackage{pdfpages}
\usepackage{wasysym}

\usepackage[a4paper, total={17cm, 24cm}, headheight=0.6cm]{geometry}
\usepackage{subcaption}  
\usepackage[pdftex,bookmarks,hidelinks]{hyperref}

\usepackage[normalem]{ulem}
\def\phaseone{Phase~I\xspace} 
\def\phasetwo{Phase~II\xspace} 

\def\expshort{DUNE\xspace}
\def\dune{\expshort}

\newcommand{\rms}{RMS\xspace} 

\newcommand{\deltacp}{\ensuremath{\delta_{\rm CP}}\xspace}   

\def\argon40{${}^{40}$Ar}       
\def\Ar39{$^{39}$Ar}
\def\Cl40{$^{40}$Cl}
\def\K40{$^{40}$K}
\def\B8{$^{8}$B}
\newcommand{\lsim}{{\;\raise0.3ex\hbox{$<$\kern-0.75em\raise-1.1ex\hbox{$\sim$}}\;}}
\newcommand{\gsim}{{\;\raise0.3ex\hbox{$>$\kern-0.75em\raise-1.1ex\hbox{$\sim$}}\;}}
\newcommand{\beq}{\begin{equation}}
\newcommand{\eeq}{\end{equation}}
\newcommand{\bea}{\begin{eqnarray}}
\newcommand{\eea}{\end{eqnarray}}

\mathchardef\minus="002D

\DeclareSIUnit \c {$c$}
\DeclareSIUnit\magn{$\times$}
\DeclareSIUnit\min{min}
\DeclareSIUnit\hr{hr}
\DeclareSIUnit\hrs{hrs}
\DeclareSIUnit\week{week}
\DeclareSIUnit\month{mo}
\DeclareSIUnit\months{mos}
\DeclareSIUnit\year{yr}
\DeclareSIUnit\years{years}
\DeclareSIUnit\yr{yr}
\DeclareSIUnit\standard{std}
\DeclareSIUnit\str{sr}
\DeclareSIUnit\ppm{ppm}
\DeclareSIUnit\ppb{ppb}
\DeclareSIUnit\ppt{ppt}
\DeclareSIUnit\pe{PE}
\DeclareSIUnit\spe{SPE}
\DeclareSIUnit\pdm{PDM}
\DeclareSIUnit\ev{events}
\DeclareSIUnit\ct{counts}
\DeclareSIUnit\neutron{\mbox{$n$}}
\DeclareSIUnit\smp{samples}
\DeclareSIUnit\Sample{S}
\DeclareSIUnit\ch{ch}
\DeclareSIUnit\hit{hit}
\DeclareSIUnit\hits{hits}
\DeclareSIUnit\bin{(\mbox{5-PE}~bin)}
\DeclareSIUnit\sgm{\mbox{$\sigma$}}
\DeclareSIUnit\rms{RMS}
\DeclareSIUnit\keVee{\mbox{keV$_{e{\rm e}}$}}
\DeclareSIUnit\keVr{\mbox{keV$_{\rm nr}$}}
\DeclareSIUnit\eVee{\mbox{eV$_{\rm ee}$}}
\DeclareSIUnit\eVr{\mbox{eV$_{\rm nr}$}}
\DeclareSIUnit\ph{photon}
\DeclareSIUnit\el{\mbox{$e^-$}}
\DeclareSIUnit\pm{\mbox{PMT}}
\DeclareSIUnit\pixel{\mbox{pixel}}
\DeclareSIUnit\inch{''}
\DeclareSIUnit\foot{'}
\DeclareSIUnit\bit{bit}
\DeclareSIUnit\sample{samples}
\DeclareSIUnit\barn{barn}
\DeclareSIUnit\bara{bar}
\DeclareSIUnit\bar{bar}
\DeclareSIUnit\barg{barg}
\DeclareSIUnit\mlardepth{\mbox(meter~of~\LAr~depth)}
\DeclareSIUnit\Curie{Ci}
\DeclareSIUnit\PSI{psi}
\DeclareSIUnit\psia{psia}
\DeclareSIUnit\atm{atm}
\DeclareSIUnit\psf{psf}
\DeclareSIUnit\pcf{pcf}
\DeclareSIUnit\parsec{pc}
\DeclareSIUnit\cps{cps}
\DeclareSIUnit\slpm{\SI{}{\liter\per\minute}}
\DeclareSIUnit\rpm{rpm}
\DeclareSIUnit\mwe{\mbox{m.w.e.}}
\DeclareSIUnit\liveday{\mbox{live-days}}
\DeclareSIUnit\days{\mbox{days}}
\DeclareSIUnit\miles{\mbox{miles}}
\DeclareSIUnit\lumens{\mbox{lm}}
\DeclareSIUnit\degreeC{\mbox{$^{\circ}$C}}
\DeclareSIUnit\degreeF{\mbox{$^{\circ}$F}}
\DeclareSIUnit\electron{\mbox{$e^-$}}
\DeclareSIUnit\Euro{\mbox{\euro}}
\DeclareSIUnit\cph{cph}
\DeclareSIUnit\neq{neq}
\DeclareSIUnit\normal{\mbox{N}}
\DeclareSIUnit\USD{\mbox{\$}}
\DeclareSIUnit\Vpercm{\mbox{V/cm}}
\DeclareSIUnit\kV{\mbox{kV}}

\DeclareSIUnit \mm {\milli\meter}
\DeclareSIUnit \cm {\centi\meter}
\DeclareSIUnit \us {\micro\second}
\DeclareSIUnit \ms {\milli\second}
\DeclareSIUnit \pA {\pico\ampere}
\DeclareSIUnit \pC {\pico\coulomb}
\DeclareSIUnit \fC {\femto\coulomb}
\DeclareSIUnit \fF {\femto\farrad}
\DeclareSIUnit \pF {\pico\farrad}
\DeclareSIUnit \mV {\milli\volt}
\DeclareSIUnit \kV {\kilo\volt}
\DeclareSIUnit \V {\volt}
\DeclareSIUnit \GOhm {\giga\ohm}
\DeclareSIUnit \MOhm {\mega\ohm}
\DeclareSIUnit \ton {\tonne}
\DeclareSIUnit \kton {\kilo\tonne}
\DeclareSIUnit \kt {\kilo\tonne}
\DeclareSIUnit \Mt {\mega\tonne}
\DeclareSIUnit \eV {\electronvolt}
\DeclareSIUnit \keV {\kilo\electronvolt}
\DeclareSIUnit \MeV {\mega\electronvolt}
\DeclareSIUnit \GeV {\giga\electronvolt}
\DeclareSIUnit \km {\kilo\meter}
\DeclareSIUnit \kW {\kilo\watt}
\DeclareSIUnit \MW {\mega\watt}
\DeclareSIUnit \MHz {\mega\hertz}
\DeclareSIUnit \kHz {\kilo\hertz}
\DeclareSIUnit \mrad {\milli\radian}
\DeclareSIUnit \year {year}
\DeclareSIUnit \POT {POT}
\DeclareSIUnit \sig {$\sigma$}
\DeclareSIUnit\parsec{pc}
\DeclareSIUnit\lightyear{ly}
\DeclareSIUnit\foot{ft}
\DeclareSIUnit\ft{ft}

\title{%
The DUNE Science Program  \\ \bigskip
\large Input to the European Strategy for Particle Physics - 2026 Update }
\author{The DUNE Collaboration
\footnote{Contact persons: Sergio Bertolucci (Sergio.Bertolucci@cern.ch), Sowjanya Gollapinni (sowjanya@lanl.gov)}
}
\date{\today}

\begin{document}

\maketitle

\begin{abstract}
    The international collaboration designing and constructing the Deep Underground Neutrino Experiment (DUNE) at the Long-Baseline Neutrino Facility (LBNF) has developed a two-phase strategy for the implementation of this leading-edge, large-scale science project. The 2023 report of the US Particle Physics Project Prioritization Panel (P5) reaffirmed this vision and strongly endorsed DUNE Phase~I and Phase~II, as did the previous European Strategy for Particle Physics. The construction of DUNE Phase~I is well underway. DUNE Phase~II consists of a third and fourth far detector module, an upgraded near detector complex, and an enhanced $>2$\,MW beam. The fourth FD module is conceived as a ``Module of Opportunity'', aimed at supporting the core DUNE science program while also expanding the physics opportunities with more advanced technologies. 

    The DUNE collaboration is submitting four main contributions to the 2026 Update of the European Strategy for Particle Physics process. This submission to the “Neutrinos and cosmic messengers”, “BSM physics” and "Dark matter and dark sector" streams focuses on the physics program of DUNE. Additional inputs related to DUNE detector technologies and R\&D, DUNE software and computing, and European contributions to Fermilab accelerator upgrades and facilities for the DUNE experiment, are also being submitted to other streams.
    
\end{abstract}

\thispagestyle{empty} 

\newpage
\pagenumbering{arabic}

\section{Scientific Context}
\label{sec:context}
\addcontentsline{toc}{section}{Executive summary}

The preponderance of matter over antimatter in the early universe, the dynamics of the supernova neutrino bursts (SNBs) that produced the heavy elements necessary for life, and the nature of physics beyond the Standard Model (BSM) are mysteries at the forefront of particle physics and astrophysics, and key to understanding the evolution of our universe.

The Deep Underground Neutrino Experiment (DUNE) will address these questions in a multidecadal science program with its world-leading liquid argon (LAr) detector technology. The international DUNE collaboration, hosted by the Fermi National Accelerator Laboratory (FNAL), is designing, developing, and constructing a near detector (ND) complex at FNAL (the near site) and a suite of four large detector modules 1300\,km downstream at the Sanford
Underground Research Facility (SURF) in South Dakota (the far site). These detectors will record neutrinos over a wide energy range, originating from a new high-intensity neutrino beamline at FNAL. The modular far detector (FD) will also detect neutrinos produced in the atmosphere and from astrophysical sources. The beamline as well as the excavations, infrastructure, and facilities for housing and supporting the DUNE detectors are provided by the Long-Baseline Neutrino Facility (LBNF).

The DUNE Collaboration was launched in 2015, following the recommendations of the 2013 update of the European Strategy for Particle Physics~\cite{2013europeanstrategy} and of the 2014 Report of the US Particle Physics Project Prioritization Panel (P5)~\cite{2014p5report}. DUNE and LBNF will complete this project in two phases, as summarized in Table~\ref{tab:phases}, based on the availability of resources and the ability to reach science milestones. The latest P5 report released in December 2023 reaffirmed this vision~\cite{2023p5report}. 

The \phaseone beamline will produce a wide-band neutrino beam with up to $1.2$~MW beam power, designed to be upgradable to $>2$\,MW. The \phaseone ND includes a moveable LArTPC with pixel readout (ND-LAr), integrated with a downstream muon spectrometer (TMS)~\cite{DUNE:2021tad}, and an on-axis magnetized neutrino detector (SAND) further downstream. The ND-LAr+TMS detector system can be moved sideways over a range of off-axis angles and neutrino energies (the DUNE-PRISM concept), for an optimal characterization of the neutrino-argon interactions. The \phaseone FD includes two LArTPC modules, each containing 17\,kt of LAr. The far detector module 1 (FD1) is a horizontal drift time projection chamber (TPC), as developed and operated in ProtoDUNE at CERN~\cite{DUNE:2020txw}. The far detector module 2 (FD2) is a vertical drift TPC~\cite{DUNE:2023nqi}. For the cryogenic infrastructure in support of the two LArTPC modules, \phaseone will include two large cryostats (one per FD module), 35\,kt of LAr, and three nitrogen refrigeration units.

\begin{table}[bp]
    \centering
    \begin{tabular}{|p{3.3cm}|p{4.7cm}|p{4.7cm}|p{2.3cm}|} \hline
        Parameter  & \phaseone      & \phasetwo & Impact \\ \hline
        FD mass    & 2 FD modules ($>20$\,kt fiducial) & 4 FD modules ($>40$\,kt fiducial LAr equivalent) & FD statistics \\  \hline
        Beam power & 1.2\,MW & $>2$\,MW   & FD statistics \\ \hline
        ND configuration  & ND-LAr+TMS, SAND & ND-LAr, ND-GAr, SAND   & Systematics \\ \hline
    \end{tabular}
    \caption{A high-level description of the two-phased approach to DUNE. The ND-LAr detector, including its capability to move sideways (DUNE-PRISM), and SAND are present in both phases of the ND. Note that the non-argon options currently under consideration for \phasetwo near and far detectors are not shown.}
    \label{tab:phases}
\end{table}

The construction of the first project phase (\phaseone), funded through commitments by a coalition of international funding agencies, is well underway. Its successful completion is currently the collaboration's highest priority. Excavation at the far site is complete, and fabrication of various beamline and detector components for \phaseone is progressing well. The facilities currently being constructed by LBNF at both the near and far sites are designed to host the full scope (\phaseone and \phasetwo) of DUNE.

\phasetwo of DUNE~\cite{DUNE:2024wvj} encompasses an enhanced multi-megawatt beam, the third and fourth FD modules, and an upgraded ND complex. The primary objective of \dune \phasetwo is a set of precise measurements of the parameters of the neutrino mixing matrix, $\theta_{23}$, $\theta_{13}$, $\Delta m^{2}_{32}$, and \deltacp, to establish Charge Conjugation-Parity Symmetry Violation (CPV) over a broad range of possible values of \deltacp, and to search for new physics in neutrino oscillations. DUNE also seeks to detect low-energy neutrinos from astrophysical sources. The additional mass brought by the \phasetwo FD modules will increase the statistics of a supernova burst signal and extend DUNE's reach beyond the Milky Way. The \phasetwo design concepts could also enable sensitive searches for new physics with solar neutrinos by lowering the detection threshold and by reducing background rates. Finally, \phasetwo will expand DUNE's new physics discovery reach for rare processes at the ND and FD sites, and for non-standard neutrino oscillations.

The \phasetwo R\&D program is a global effort with contributions from all DUNE partners and potential new collaborators.  Part of the R\&D described in this document is carried out within the framework of the CERN detector R\&D collaborations and those being formed under the umbrella of the Coordinating Panel for Advanced Detectors (CPAD) in the US. 

Support by the European science community is critical to the success of DUNE's ground-breaking neutrino science program. A $39\%$ of DUNE collaborators are based at European universities and research institutions. They provide key contributions to the Far and Near detector components, computing, as well as to the LBNF beam. These contributions, funded largely by European national funding agencies, 
are key to the design and construction of LBNF and DUNE Phase I. The R\&D and validation of the detector technologies for DUNE's first two far detector modules has been enabled by the Neutrino Platform at CERN through its ProtoDUNE program, which has been an extraordinary success. 
Finally, CERN provides key LBNF infrastructure, in particular through the
procurement of the FD cryostats. CERN also plays an
important role as European hub for the DUNE Collaboration, regularly hosting collaboration meetings
and workshops. 

DUNE relies on CERN's continued support, specifically by:

\begin{itemize}
    \item providing infrastructure, technical expertise, resources, support for DUNE R\&D, and detector technology and physics validation through the Neutrino Platform and the ProtoDUNE facilities;
    \item additional support of the far site facility (LBNF) through the procurement of cryostats for the third and fourth modules;
    \item continued technical and engineering support during the DUNE construction phase;
    \item CERN support for addressing growing DUNE computing needs;
    \item support and coordination of the DRD collaborations;
    \item providing important physics and technical leadership on DUNE through CERN's EP-NU group and CERN’s Neutrino Platform;
    \item providing facilities for GeV-scale beam tests for detector R\&D at CERN, and by supporting dedicated experiments to measure hadron production cross-sections for the DUNE environment.
\end{itemize}

\section{Objectives}
\label{sec:physics}

\begin{table}[tbp]
\centering
\begin{tabular}{|p{6.cm}|p{6.0cm}|p{3.5cm}|} \hline

{\bf Benchmark} & {\bf DUNE's Projected Reach} & {\bf Assumptions} \\ \hline

Sensitivity to CP violation & 5$\sigma$ (50\% of $\deltacp$ values) & 600~kt-MW-yr \\ 
 & 3$\sigma$ (75\% of $\deltacp$ values) & 1000~kt-MW-yr \\ \hline
Precision on $\deltacp$ as a function of true $\deltacp$ & 11$^{\circ}$ ($\deltacp = 0$) -- 27$^{\circ}$ ($\deltacp =-\pi/2$) & 300~kt-MW-yr \\
 & 8$^{\circ}$ ($\deltacp = 0$) -- 22$^{\circ}$ ($\deltacp =-\pi/2$) & 600~kt-MW-yr \\
  & 7$^{\circ}$ ($\deltacp = 0$) -- 18$^{\circ}$ ($\deltacp =-\pi/2$) & 1000~kt-MW-yr \\ \hline
Sensitivity to mass ordering & 5$\sigma$ (100\% of $\deltacp$ values) & 70~kt-MW-yr \\ \hline
Precision on mixing angles and mass differences in PMNS & sin$^22\theta_{13}$ resolution of 5\% & 1000~kt-MW-yr \\ 
 & $\Delta m^2_{32}$ resolution of 1\% & 100~kt-MW-yr \\
  & $\Delta m^2_{32}$ resolution of 0.4\% & 1000~kt-MW-yr \\ \hline\hline
Sensitivity to light sterile neutrinos, (3+1) model & $\sin^22\theta_{14}> 2\times 10^{-3}$ at 90\% C.L. & 300~kt-MW-yr \\
 & $\sin^22\theta_{24}> 2\times 10^{-4}$ at 90\% C.L. & 300~kt-MW-yr \\ \hline
Sensitivity to non-standard neutrino interactions affecting propagation & $\lvert\epsilon^m_{e\mu}\rvert>0.1$ at 90\% C.L. & 300~kt-MW-yr \\
 & $\lvert\epsilon^m_{e\tau}\rvert>0.3$ at 90\% C.L. & 300~kt-MW-yr \\ \hline
 Sensitivity to heavy neutral lepton mixing & $\lvert U_{eN}\rvert^2,\ \lvert U_{\mu N}\rvert^2>2\times 10^{-10}$ at 90\% C.L. & $8\times 10^{21}$ POT, no bgr \\
  & $\lvert U_{\tau N}\rvert^2>2\times 10^{-7}$ at 90\% C.L. & $8\times 10^{21}$ POT, no bgr \\ \hline
Sensitivity to light dark matter ($\alpha_{D}=0.5, M_{V}=3M_{\chi,\phi}$) & $Y \equiv \varepsilon^{2}\alpha_{D}\left(\frac{M_{\chi,\phi}}{M_{V}}\right)^{4}   \gtrsim  3\times10^{-11}$ at $m_\chi = 10$ MeV & 3.5 yr, on-axis ND \\
 & $Y \gtrsim 1\times10^{-11} $ at $m_{\chi}=10$ MeV & 3.5 yr, 24~m off-axis ND \\ \hline
 Sensitivity to proton decay & $<1.3\times 10^{34}$ years for $p\to K^+\overline{\nu}$ partial lifetime at 90\% C.L. & 400~kt-yr \\ \hline \hline
Sensitivity to supernova neutrinos & 3,300 $\nu_e + ^{40}Ar \to e^- + ^{40}K^{\ast}$ events & 40~kt of LAr, 10~kpc SN distance \\ \hline
Pointing capabilities of neutrino experiments to identify sources & 4.3$^{\circ}$ SN pointing resolution & 40~kt of LAr, 10~kpc SN distance \\ 
 \hline

\end{tabular}
\caption{DUNE's projected reach for selected PPG benchmark measurements for the "Neutrinos and cosmic messengers", "BSM physics" and "Dark matter and dark sector" themes. For BSM search entries, DUNE's projected reach refers to regions of parameter space where DUNE could discover BSM physics. From~\cite{DUNE:2022aul,DUNE:2020jqi,DUNE:2020fgq,DUNE:2020zfm,DUNE:2024ptd}.}
\label{tab:physics_benchmarks}
\end{table}

DUNE provides guaranteed high-profile discoveries and precision measurements, on a short timescale by HEP standards, through its long-baseline neutrino oscillation physics program. At the same time, it provides compelling neutrino astrophysics and BSM physics reach, the latter both through direct searches and through tests of the three-flavor paradigm in neutrino oscillations. This section discusses selected DUNE science objectives in all these areas, highlighting the benefits of the \phasetwo beam and detectors. DUNE’s projected reach for selected PPG benchmark measurements for the ``Neutrinos and cosmic messengers'', "BSM physics" and "Dark matter and dark sector" themes is reported in Table~\ref{tab:physics_benchmarks}.

\subsection{Long-baseline neutrino oscillation physics}
\label{subsec:physics_lbl}

DUNE is designed to measure the appearance of electron (anti)neutrinos ($\nu_{e}$ or $\bar{\nu}_{e}$) and the disappearance of muon (anti)neutrinos ($\nu_{\mu}$ or $\bar{\nu}_{\mu}$) as functions of neutrino energy in a wide-band beam. DUNE is sensitive to all the parameters governing $\nu_{1}-\nu_{3}$ and $\nu_{2}-\nu_{3}$ mixing in the three-flavor model simultaneously: $\theta_{23}$, $\theta_{13}$, $\Delta m^{2}_{32}$ (including its sign, which specifies the neutrino mass ordering), and the CP violating phase \deltacp. 

The goals of the oscillation physics program of \phasetwo are to make high-precision measurements of all four parameters, to establish CPV at high significance over a broad range of possible values of \deltacp, and to test the three-flavor paradigm as a way to search for new physics in neutrino oscillations. Achieving these goals requires $600-1000$\,kt$\cdot$MW$\cdot$yr of data statistics, depending on the measurement. This can be achieved by operating for $6-10$ additional calendar years with a greater than $2$\,MW beam and a FD of $40$\,kt LAr equivalent fiducial mass. Without Phase II, the additional time required would be $24-40$ years. 

In addition to the high statistics enabled by the increased mass, \phasetwo oscillation physics goals also require systematic uncertainties to be constrained at the few percent level, which is beyond what is achieved in current experiments. This is the role of the \phasetwo ND; the main improvement is the addition of ND-GAr, a magnetized high-pressure gaseous argon time projection chamber surrounded by an electromagnetic calorimeter and by a muon detector, that will measure neutrino-argon interactions with unprecedented precision.

With \phasetwo, DUNE will measure \deltacp with world-leading precision. The wide-band beam enables a broad spectral measurement that is crucial for resolving degeneracies between different values of \deltacp, and between \deltacp and $\theta_{23}$. DUNE will measure $\theta_{23}$ with world-leading precision and determine the octant if it is sufficiently non-maximal. The measurements of $\theta_{13}$ and $\Delta m^{2}_{32}$ are highly complementary with the current $\theta_{13}$ measurement from Daya Bay~\cite{DayaBay:2022orm} and the planned $\Delta m^{2}_{32}$ measurement from JUNO~\cite{JUNO:2022mxj}, respectively, which are performed with a different neutrino flavor. DUNE is also highly complementary with Hyper-K~\cite{Hyper-Kamiokande:2018ofw}, which measures the same flavor transitions at a similar $L/E$ but with a much smaller matter effect, lower neutrino energy, narrower energy spectrum, and an entirely different detector technology that will yield different systematics. Comparing the results obtained over this wide range of conditions will provide a more complete and robust test of the three-flavor model.

DUNE is also sensitive to BSM physics that modifies neutrino oscillations, including non-unitary mixing, non-standard interactions, CPT violation, and the possible existence of additional neutrino species (see Section~\ref{subsec:physics_bsm}). 

{\bf The increased statistics enabled by doubling the FD mass and beam intensity, and the reduction in systematics enabled by the \phasetwo ND, are critical to the precision long-baseline program.}


\subsection{Neutrino astrophysics, supernova, and solar neutrinos}
\label{subsec:physics_astro}

DUNE will detect MeV-scale neutrinos from astrophysical sources, primarily from the sun and from supernova explosions. With argon as the target, DUNE is sensitive to the astrophysical $\nu_e$ flux for energies below 100\,MeV and above 5\,MeV due to the relatively large cross-section for the process: $\nu_e+^{40}\mathrm{Ar}\rightarrow e^-+^{40}\mathrm{K}^\ast$. DUNE's $\nu_e$ sensitivity is unique, and highly complementary to existing and proposed experiments aiming for similar astrophysical neutrino measurements.

In the context of a core-collapse supernova (CCSN), the $\nu_e$ sensitivity is most striking immediately following the SNB, when $\nu_e$ emission from neutronization in the stellar core dominates. As the neutrinos escape a CCSN before the first optical signal, pin-pointing the source of the SNB is crucial to facilitate optical observation of the initial stages of the supernova. DUNE is also sensitive to fundamental neutrino properties via SNB detection, including the neutrino mass ordering~\cite{Dasgupta:2007ws}, neutrino self-interaction strength~\cite{Chang:2022aas}, and the absolute neutrino mass via time of flight from the supernova to Earth~\cite{Pompa:2022cxc}. 

DUNE will also measure $^8$B solar neutrinos and improve upon current solar measurements of $\Delta m^2_{21}$ via the day-night asymmetry induced by Earth matter effects. DUNE will also make the first observation of the ``hep" flux produced via the $^3$He + p $\to$ $^4$He + e$^+$ + $\nu_e$ nuclear fusion, exceeding $5\sigma$ even in \phaseone.

{\bf The additional target mass of \phasetwo will enhance DUNE's MeV-scale physics sensitivity.} The SNB trigger efficiency, reconstruction of the supernova direction, and precise measurement of the supernova spectral parameters are driven by the number of neutrino interactions~\cite{DUNE:2020zfm}. The day-night asymmetry measurement is also statistically limited and benefits from additional mass.

{\bf Beyond the additional target mass, improved technologies with \phasetwo can greatly expand the low-energy physics reach of DUNE.} For solar neutrinos, improved technologies could reduce the threshold from $\sim 10$ MeV to $\sim 5$ MeV, and the energy resolution from $(10-20)\%$ to $\sim 2\%$, which would substantially expand the physics reach. DUNE would improve the determination of solar neutrino oscillation parameters and would probe the upturn in the $\nu_e$ survival probability, which is the transition between the low-energy regime where vacuum oscillations dominate and the high-energy regime where the oscillation probability is determined by Mikheyev-Smirnov-Wolfenstein (MSW) matter effects inside the sun. Measurements of carbon nitrogen oxygen (CNO) solar neutrinos that could distinguish between solar metallicity models~\cite{Bezerra:2023gvl} would also be possible. 

With increased photodetector coverage and use of $^{39}$Ar-depleted argon, a DUNE LArTPC module would be sensitive to faint light flashes from Coherent Elastic Neutrino-Nucleus Scattering (CE$\nu$NS) during a SNB~\cite{Bezerra:2023gvl}. As a neutral current (NC) process, this channel is sensitive to all flavors, complementary to the $\nu_e$ charged current (CC) signal. 

DUNE's low-energy physics can reach beyond astrophysical neutrinos in \phasetwo. With shielding, heavy fiducialization, improved energy resolution and depleted argon, a DUNE FD module could perform a competitive weakly-interacting massive particle (WIMP) dark matter search complementary to future argon dark matter experiments~\cite{Church:2020env}. By introducing a large mass fraction of $^{136}$Xe to a FD module, DUNE could also perform neutrinoless double-$\beta$ decay (0$\nu\beta\beta$) searches.

A water-based liquid scintillator module (e.g., \textsc{Theia}~\cite{Theia:2019non}) would sacrifice part of the SNB $\nu_e$ statistics for a significant increase in $\bar{\nu}_e$ events through inverse beta decay, with a threshold of $\sim$2~MeV. Such a module would detect approximately 5000 $\bar{\nu}_e$ interactions from a SNB at 10\,kpc distance. The scintillation light would provide a tag for neutrons to allow separation between inverse $\beta$ decay events and directionally-sensitive elastic scattering reactions. Using the high light output from the scintillation, such a module would also be sensitive to pre-supernova neutrinos~\cite{Mukhopadhyay_2020}, alerting neutrino experiments to an upcoming SNB. A \textsc{Theia} module would also observe geo-neutrinos and reactor neutrinos~\cite{Theia:2019non}, and $^{130}$Te isotope loading could enable sensitivity to 0$\nu\beta\beta$.

\subsection{Physics beyond the Standard Model}
\label{subsec:physics_bsm}

DUNE has discovery sensitivity to a diverse range of BSM physics in a relatively unexplored phase space, complementary to searches at collider experiments and other precision experiments. BSM physics accessible at DUNE can be divided into three major areas: rare processes in the beam observed at the ND, rare event searches at the FD, and BSM neutrino oscillations. 

The high intensity and high energy of the LBNF proton beam enables DUNE to search for long-lived exotic particles that are produced in the target and either scatter or decay in the ND. This opens up opportunities for DUNE to explore various BSM particles, including Light Dark Matter (LDM), Heavy Neutral Leptons (HNLs), and Axion-Like Particles (ALPs). LDMs can be the decay products of dark photons produced through kinetic mixing with the SM photons created in the LBNF neutrino target. For LDM, it has been shown that the use of the DUNE-PRISM concept \cite{DeRomeri:2019kic} and an opportunistic beam dump mode \cite{Brdar:2022vum} in ND could significantly improve the sensitivity. The \phasetwo beam upgrade is expected to further improve this. For HNLs, the sensitivity, assuming zero background, is expected to be world-leading at masses below the tau mass, complementing the LHC searches for heavier mass particles~\cite{Coloma:2020lgy}. For ALPs, DUNE is expected to improve existing constraints, particularly for ALP masses below the kaon mass~\cite{Kelly:2020dda,Coloma:2023oxx}. For a wide range of ALP lifetimes, the sensitivity on the product of the ALP production and decay branching ratios would improve by many orders of magnitude. Another rare event search at DUNE is trident production, where neutrino scattering on a heavy nucleus produces a pair of oppositely charged leptons. Trident production serves as a powerful probe of BSM physics in the leptonic sector~\cite{Altmannshofer:2014pba,Ballett:2019xoj}.

The main upgrade to the ND in \phasetwo is ND-GAr. \textbf{ND-GAr vastly expands DUNE's BSM physics reach}, as the signal scales with detector volume, while the background from SM neutrino interactions scales with mass. ND-GAr's excellent vertex reconstruction further suppresses SM background, particularly nuclear breakup, which does not occur in decay signals. This makes ND-GAr an ideal detector for HNL and ALP searches. Additionally, ND-GAr plays a crucial role in the search for trident production, particularly for dimuon tridents ($\nu_{\mu} \to \nu_{\mu} \mu^{+} \mu^{-}$ and $\nu_{\bar{\mu}} \to \nu_{\bar{\mu}} \mu^{+} \mu^{-}$). While the signal rate is lower, ND-GAr offers much better background rejection compared to the \phaseone ND, allowing for more efficient separation of the signal from the background.

{\bf \phasetwo will also enhance BSM rare event searches at the FD}, in particular searches that are expected to be nearly background-free at the scale of the experiment's full exposure. In such cases, the decay or scattering rate sensitivity will be inversely proportional to the FD exposure (in kt$\cdot$yr), and added exposure in \phasetwo FD modules will be significant. Background-free (or quasi-background-free) searches at the FD may include a number of baryon-number-violating nucleon decay modes~\cite{DUNE:2020fgq}. 

{\bf DUNE is sensitive to neutrino oscillation scenarios beyond the standard three-flavor picture}, including sterile neutrinos, non-standard interactions (NSI), and PMNS non-unitarity. DUNE is particularly sensitive to NSI matter effects due to the long baseline. These searches rely on both the ND and FD, and require high precision and very large exposures, such that both the \phasetwo ND and FD are important. DUNE has substantial flux above the $\tau$ appearance threshold, and is uniquely sensitive to $\tau$ appearance, which adds additional sensitivity to non-unitarity and NSI~\cite{DeGouvea:2019kea,Ghoshal:2019pab}. DUNE will also search for anomalous $\tau$ appearance in the ND, which is enhanced by ND-GAr in \phasetwo due to its superior reconstruction of high-energy muons.

\section{Methodology}
\label{sec:methodology}

\begin{table}[tbp]
\centering
\begin{tabular}{|p{3.cm}|p{9.5cm}|}\hline
\multicolumn{1}{|c|}{Technology} & \multicolumn{1}{|c|}{Description}\\ \hline 
APEX & Large-area photon detection system on TPC field cage  \\ \hline

CRP & Projective charge readout using strips \\ \hline

LArPix, Q-Pix & 3D charge readout solutions using charge pixels \\ \hline

SoLAr, Q-Pix-LiLAr, LightPix & Integrated charge and light readout pixel solutions on anode \\ \hline

ARIADNE & Dual-phase LAr TPC with 3D optical readout of charge signal  \\ \hline

\textsc{Theia} & Water-based, hybrid Cherenkov plus scintillation, detection concept \\ \hline
\end{tabular}
\caption{Detector technologies currently being considered for the \phasetwo FD modules~\cite{DUNE:2024wvj}.} 
\label{tab:fd3_fd4_options}
\end{table}

While several options are under consideration for the \phasetwo components of the far and near site detectors, key elements have already been defined.

{\bf Two additional FD modules, FD3 and FD4, will be added at the far site, for a total of four.} The DUNE FD2 vertical drift (VD) technology forms the basis for the envisioned design of FD3. Tab.~\ref{tab:fd3_fd4_options} summarizes new technologies being considered for \phasetwo. For FD3, further optimization of the strip-based charge readout planes (CRPs) could improve the charge readout. Improved photon detection is also being pursued, for example with aluminum profiles with embedded X-ARAPUCA (APEX), which integrates much larger area photon detectors into the field cage. More ambitious module concepts are under consideration for FD4. Pixel-based readout is the baseline design in ND-LAr and an option for an FD module. Two readout schemes are under study, LArPix (used by ND-LAr) and Q-Pix. The pixel anode could also be designed to integrate charge and light readout (the SoLAr concept, which could improve solar neutrino reach). A dual-phase module with optical readout (ARIADNE) could be implemented as either a single drift volume, or combined with another single-phase readout scheme for the bottom drift region. Beyond improvements to readout, the potential to enhance the physics scope of DUNE \phasetwo with lower energy thresholds relies on greater control of radioactive backgrounds, including shielding against external neutrons and careful detector material selection programs using radioactive assay techniques. More ambitious ideas to drastically reduce backgrounds with depleted argon, targeting very low energy physics, are also being considered.
A non-argon option is also under consideration as an alternative technology for FD4; a \textsc{Theia} module would combine Cherenkov and scintillation detection with a water-based scintillator.

{\bf The main improvement of the \phasetwo Near Detector is the addition of ND-GAr.} ND-GAr will serve both as a new muon spectrometer for ND-LAr, replacing the TMS in this capacity, and as a new neutrino detector to study neutrino-argon interactions occurring in the high-pressure gaseous argon TPC. In addition, upgrades to the ND-LAr and SAND systems are considered, as well as potential ND options in the case of a non-argon technology for FD4.
    
{\bf A beam upgrade to increase the intensity to $>$2 MW}. This is achieved by ACE-MIRT, which increases the frequency of beam spills by nearly a factor of two. While this is part of \phasetwo, it is possible to implement the upgrades that comprise ACE-MIRT before DUNE beam data taking begins.


\section{Readiness, expected challenges, and timeline}

All DUNE \phasetwo FD detector technologies listed in Tab.~\ref{tab:fd3_fd4_options} are technically mature. Only a limited number of key R\&D goals remain to be demonstrated for each technology, with currently achieved Technology Readiness Levels (TRLs) for those estimated to be 3 or more\footnote{We adopt the TRL definition as per \cite{trl}.}. Large-scale prototyping plans have been defined for each technology~\cite{DUNE:2024wvj}. The ProtoDUNEs at CERN will continue to serve as essential platforms to demonstrate several of these technologies and their potential for integration. 

The reference designs for the Phase~II modules FD3 and FD4 are improved versions of the vertical drift LArTPC technology, with \textsc{Theia} serving as an alternative technology choice for FD4. Decisions on the technology choices for FD3 and FD4 are expected to come no later than 2027 and 2028, respectively. 
 
The final design milestones for \phasetwo FD modules are driven by the number and extent of the improvements planned. For example, in the case of a FD2-like module where the only improvements are optimization of CRPs and the APEX light system, we can envision being ready for a final design by 2028 in a technically limited schedule. In this scenario, the earliest start for installation of FD3 can be anticipated in 2029 with completion of installation and filling in 2034. Alternatively, if we were to implement pixel-based technologies such as LArPix, Q-Pix, SoLAr, or ARIADNE (top anode plane only), the final design milestone would likely be delayed until at least 2030. An asymmetric dual-phase VD LArTPC for FD4, with a single drift volume instrumented via a single ARIADNE readout plane, would require significant changes to the high voltage system. It is possible to reach the final design milestone for this option by 2031-32. In the case of the \textsc{Theia} option, a final design milestone no earlier than 2033 is anticipated. 

For the \phasetwo ND, the detector requirements are in the process of being defined, with an aim toward a complete conceptual design in the late 2020s. A final design milestone would follow in the early 2030s, with start of operations expected in the mid-2030s. To this end, the critical R\&D items to demonstrate the ND-GAr feasibility have been identified. Those include R\&D on superconducting cables for the magnet, testing of the full TPC readout chain from amplification technology to readout electronics at high pressure, and scintillation light readout. Several beam tests have been completed, providing important lessons learned. Further beam tests are being planned, with CERN being a potential site. The ND-LAr and SAND components of the ND may also be upgraded for \phasetwo. A decision on these possible upgrade paths will come after the \phaseone ND is commissioned.

\section{Summary}
\label{sec:summary}

DUNE \phaseone is well underway, with key contributions from Europe. In particular, the CERN Neutrino Platform has enabled critical R\&D and validation of the detector technologies for the first two far detectors of DUNE with great success. DUNE relies on continued support from Europe to successfully complete \phaseone and realize \phasetwo. Key requests from DUNE for European Strategy include maintaining ProtoDUNEs at the CERN Neutrino Platform, supporting LBNF infrastructure for DUNE \phasetwo with cryostats for third and fourth far detector modules, supporting European contributions to the \phasetwo near and far detectors and upgraded beamline, and continuing to support scientific and technical leadership in DUNE from Europe. In partnership with Europe, DUNE aims to complete the full scope of DUNE, enabling a multi-decadal program of groundbreaking science with neutrinos.

\newpage



\bibliographystyle{utphys} 
\bibliography{common/references} 

\end{document}